\documentclass[a4paper, 11pt]{article}
\usepackage[margin=25mm]{geometry}
\usepackage{booktabs}
\usepackage{multirow}
\usepackage{footnote}
\usepackage[shortlabels]{enumitem}
\usepackage{float}
\usepackage{comment}
\usepackage{amsmath}
\usepackage{amsfonts}
\usepackage{titlesec}
\usepackage{bm}
\usepackage{lscape} 
\usepackage{graphicx,comment}
\usepackage{float}
\usepackage{url}
\usepackage{color}
\usepackage{xargs, xstring}
\usepackage{subfig}
\usepackage{verbatim}
\usepackage{setspace} 
\usepackage[round]{natbib}

\providecommand{\keywords}[1]
{
  \small	
  \textbf{\textit{Keywords---}} #1
}

\newcommand{\be}{\begin{eqnarray}}
\newcommand{\ee}{\end{eqnarray}}
\newcommand{\bee}{\begin{eqnarray*}}
\newcommand{\eee}{\end{eqnarray*}}
\newcommand{\bi}{\begin{enumerate}}
\newcommand{\ei}{\end{enumerate}}

\newcommand{\sumi}{\sum_{i=1}^n}

\newcommand{\logit}{\mbox{logit}}

\newcommand{\OW}{\mbox{\tiny{OW}}}

\newcommand{\DIF}{\mbox{\tiny{UNADJ}}}

\newcommand{\imp}{\mbox{\tiny{imp}}}

\usepackage[noabbrev,capitalize]{cleveref}
\crefformat{equation}{(#2#1#3)}

\newtheorem{prop}{Proposition}
\newtheorem{assumption}{Assumption}

\title{Covariate Adjustment in Randomized Clinical Trials with Missing Covariate and Outcome Data}
\author{Chia-Rui Chang$^{1}$, Yue Song$^{1}$, Fan Li$^{2}$, Rui Wang$^{1,3}$  \\
        \\
        \small $^{1}$Department of Biostatistics, Harvard T. H. Chan School of Public Health, Massachusetts, USA \\
        \small $^{2}$Department of Statistical Science, Duke University, North Carolina, USA \\
        \small $^{2}$Department of Population Medicine, Harvard Pilgrim Health Care Institute and Harvard Medical School, \\
        \small Massachusetts, USA \\
}
\date{} 

\begin{document}
\maketitle

\doublespacing
\begin{abstract}
\noindent When analyzing data from randomized clinical trials, covariate adjustment can be used to account for chance imbalance in baseline covariates and to increase precision of the treatment effect estimate. A practical barrier to covariate adjustment is the presence of missing data. In this paper, in the light of recent theoretical advancement, we first review several covariate adjustment methods with incomplete covariate data. We investigate the implications of the missing data mechanism on estimating the average treatment effect in randomized clinical trials with continuous or binary outcomes. In parallel, we consider settings where the outcome data are fully observed or are missing at random; in the latter setting, we propose a full weighting approach that combines inverse probability weighting for adjusting missing outcomes and overlap weighting for covariate adjustment. We highlight the importance of including the interaction terms between the missingness indicators and covariates as predictors in the models. We conduct comprehensive simulation studies to examine the finite-sample performance of the proposed methods and compare with a range of common alternatives. We find that conducting the proposed adjustment methods generally improves the precision of treatment effect estimates regardless of the imputation methods when the adjusted covariate is associated with the outcome. We apply the methods to the Childhood Adenotonsillectomy Trial to assess the effect of adenotonsillectomy on neurocognitive functioning scores.
\end{abstract} \hspace{10pt}

\noindent 

\section{Introduction}
\label{sec:intro}
Randomized controlled trials (RCT) are the gold standard for evaluating the efficacy and safety of new treatments. Randomization ensures both measured and unmeasured covariates are balanced in large samples. However, chance imbalance of patient baseline characteristics may occur due to the random nature in allocating the treatment \citep{Senn1989,Ciolino2015}, especially when the sample size is small \citep{Thompson2015}. Adjusting for imbalanced covariates, particularly the ones that are predictive of the outcome, i.e. prognostic risk factors, in analysis can increase precision in estimating the treatment effects as well as improve face validity  \citep{Ciolino2015,Pocock2002, lin2013agnostic, benkeser2021improving}. One common method for covariate adjustment in RCT is regression adjustment \citep{yang2001efficiency,Kahan2016,leon2003semiparametric,tsiatis2008covariate,Zhang2008}, where the outcome is regressed on the treatment and covariates \citep{tsiatis2008covariate, lin2013agnostic}, and the treatment effect is estimated by the coefficient of the treatment variable. This is also known as the analysis of covariance (ANCOVA) method.
Recently, propensity score weighting  has been proposed as an alternative covariate adjustment method with several conceptual and practical advantages \citep{williamson2014variance,shen2014inverse,Colantuoni2015,zeng2021propensity}. The two methods are shown to be asymptotically equivalent \citep{williamson2014variance,zeng2021propensity} and empirical analyses also suggest similar finite sample performances in many situations. 

A practical barrier to covariate adjustment is the presence of missing data. When covariates are partially observed, the analysts have several choices of estimating the treatment effect: \emph{(i)} the \emph{unadjusted} analysis, where one obtains a simple difference-in-mean estimator without using any covariate information; \emph{(ii)} the \emph{complete-case} analysis, where only \emph{units} with fully observed covariates are used in the adjustment; \emph{(iii)} the \emph{complete-covariate} analysis, where only fully observed \emph{covariates} are used in the adjustment; \emph{(iv)} the \emph{missing-indicator} method \citep{groenwold2012missing}, where one includes a dummy variable for missingness for the partially observed covariate in regression adjustment, and all missing values are set to the same value; \emph{(v)} the \emph{imputation} method, where one first imputes the missing covariate values and then conducts adjustment on the completed data. A consensus in the literature is that the \emph{complete-case} analysis should be avoided \citep{schemper1990efficient,white2005adjusting,groenwold2012missing,sullivan2018should, kayembe2020imputation,kayembe2022imputation}. Moreover, \cite{white2005adjusting} showed that the \emph{missing-indicator} method generally performs well regardless of whether missingness of baseline covariates is predictive of the outcome, whereas \emph{mean imputation} is adequate when missingness is completely random (MCAR). \cite{groenwold2012missing} clarified that the \emph{missing-indicator} method is valid for randomized experiments and obeys the intention-to-treat principle, but it can lead to bias in observational studies, which was also pointed out earlier \citep{greenland1995critical}. \cite{sullivan2018should} and \cite{kayembe2020imputation} investigated the use of the popular multiple imputation (MI) \citep{Rubin1987} method in regression adjustment; both concluded that MI does not improve over simpler imputation methods in most cases.

An important recent theoretical development regarding regression adjustment with missing covariates is due to \cite{zhao2022adjust}. They proposed a modified missing-indicator method, where one imputes the missing covariates with zeros and then applies Lin's regression estimator \citep{lin2013agnostic}, which postulates an outcome model including all the imputed covariates, the missing indicators, and the covariate-treatment interactions. They showed that this method is asymptotically more efficient than the \emph{unadjusted} estimator, regardless of the missing data mechanism. They also showed theoretically that the choice of imputation method and the validity of the imputation model matters little, that is, simple imputation by a constant leads to the same precision gain as imputations based on correctly specified models. 

Despite the above advancements, covariate adjustment with missing data still has not been widely adopted in practice. And a few important questions remain. First, propensity score weighting has been advocated as a outcome-model-free and asymptotically equivalent alternative to regression adjustment, but it is unclear how to accommodate partially missing covariates in this approach. Second, more importantly, most clinical trials have missing data in both covariates and outcomes. Despite the large literature on the analysis of missing outcome data in randomized trials, there has been little investigation on the case of missingness in both covariates and outcomes. When there is missing data in the outcomes, complete case analysis that ignores units with missing outcomes can lead to biased treatment effect estimates as well as efficiency loss. Under the missing at random (MAR) assumption \citep{rubin1976inference}, two commonly used approaches for handling missing outcomes are: \emph{(i)} the \emph{imputation} method  \citep{rubin1996multiple}, where one first imputes the missing outcomes based on a model and then estimate the treatment effects; \emph{(ii)} the \emph{inverse probability weighting} (IPW) method \citep{horvitz1952generalization,li2013weighting}, where one estimates the probability of missingness of each unit given the observed variables and weighs the outcome by the inverse of that probability in estimating the treatment effect \citep{seaman2013review}. Because one cannot verify the correctness of the MAR assumption using the observed data, sensitivity analyses such as those based on pattern-mixture and selection models are recommended to evaluate the robustness of conclusions under alternative 
assumptions about the missingness process \citep{little2012prevention}. It is important to note that consideration of missing data is closely tied with the target estimand. The focus of this paper is on handling missing covariates in estimating treatment effects with covariate adjustments. Recognizing that outcome missingness is common in practice, we consider both the settings when the outcome data are complete or are partially observed, in the latter case, we will assume an MAR mechanism and adopt an inverse probability weighted approach.  

The rest of the paper is organized as follows. We first briefly review covariate adjustment with complete data (Section \ref{sec:CovAdj}). We then discuss covariate adjustment with (i) missingness only in covariates (Section \ref{sec:missingX}) and (ii) missingness in both covariates and outcome (Section \ref{sec:missingXY}). We specify two versions of missing at random assumptions of the outcome missingness mechanisms and provide corresponding models to estimate the outcome missingness probability (i.e., the propensity) . We further propose a full weighting approach that combines inverse probability weighting for adjusting missing outcomes and overlap weighting for covariate adjustment \citep{zeng2021propensity}. 
We then conduct extensive simulation studies to compare the proposed methods with a range of alternatives (Section \ref{sec:simulations}). In Section \ref{sec:application}, we apply the methods to the Childhood Adenotonsillectomy Trial (CHAT) \citep{marcus2013randomized} and illustrate one way to assess how the magnitude of potential efficiency gain vary according to the proportion of missingness and the covariate's association with the outcome. 

\section{Covariate adjustment with complete data} \label{sec:CovAdj}
Consider a randomized trial with two arms and $N$ patients, where $N_1$ and $N_0$ patients are randomized into the treatment and control arm, respectively. Let $Z_{i}=z$ be the binary treatment indicator, with $z=1$ indicates treatment and $z=0$ control. Under the potential outcome framework \citep{Neyman1923} and the stable unit treatment value assumption (SUTVA), each unit has two potential outcomes $\{Y_{i}(1),Y_{i}(0)\}$, corresponding to the treatment and the control conditions, respectively. For each unit, only the potential outcome under the actual assigned condition is observed, denoted by $Y_i=Z_iY_i(1)+(1-Z_i)Y_i(0)$. A collection of $p$ baseline variables are recorded for each unit, denoted by $X_i=( X_{i1},\ldots,X_{ip})^T$. 
A common causal estimand is the average treatment effect (ATE):
\begin{equation}
\label{def:ATE}
\tau=E\{Y_{i}(1)-Y_{i}(0)\}
\end{equation}
In randomized trials, the treatment $Z$ is randomly assigned with a fixed probability $r$, that is $\Pr(Z_i=1|X_i,Y_i(1),Y_i(0))=\Pr(Z_i=1)=r$. A  common study design uses balanced assignment with $r={1}/{2}$. Under randomization, we have $\tau=E(Y_i|Z_i=1)-E(Y_i|Z_i=0)$, and thus an unbiased estimator of the ATE is the unadjusted difference-in-means estimator: $\hat{\tau}^{\DIF}={\sum_{i=1}^{N}Z_{i}Y_{i}}/{\sumi Z_i}-{\sum_{i=1}^{N}(1-Z_{i})Y_{i}}/{\sumi (1-Z_i)}.$ Previous research has shown that adjusting for chance imbalance in covariates that are predictive of the outcome, namely prognostic covariates, can substantially improve statistical efficiency over the unadjusted estimator.  

There are two classes of covariate adjustment methods in randomized trials. The first class is outcome regression \citep{lin2013agnostic, yang2001efficiency,Kahan2016,leon2003semiparametric,tsiatis2008covariate,Zhang2008}---also known ordinary least square (OLS) regression---based on the analysis of covariance (ANCOVA) models, where the outcome is regressed on the treatment and covariates and the OLS estimator, $\tau^{ols}$, is estimated by the coefficient of the treatment variable \citep{yang2001efficiency,leon2003semiparametric,tsiatis2008covariate}. \cite{lin2013agnostic} showed that for efficient outcome regression adjustment it is critical to include the full set of covariate-treatment interaction terms in the OLS model:
\begin{equation} \label{eq:OLS-lin}
    Y_i \sim Z_i + X_i + Z_iX_i. 
\end{equation}
When the randomization probability is $1/2$, OLS returns consistent point and interval estimates even if the outcome model is misspecified \citep{yang2001efficiency,lin2013agnostic,wang2019analysis}. However, misspecification of the outcome model can decrease precision in unbalanced experiments with treatment effect heterogeneity \citep{Freedman2008}. Despite the asymptotic argument, including a full set of covariate-treatment interaction terms sometimes leads to higher variance in small samples. OLS models can also lead to unstable estimates when applied to rare binary outcomes \citep{zeng2021propensity}.

The second class of covariate adjustment methods is propensity score weighting \citep{williamson2014variance,shen2014inverse,Colantuoni2015, zeng2021propensity}. The general form of a weighted estimator for the ATE is
\begin{equation}
\label{eq:sampleWATE}
\hat{\tau}^w=\frac{\sumi w_1(x_i)Z_i Y_i}{\sumi w_1(x_i)Z_i} -
              \frac{\sumi w_0(x_i)(1-Z_i) Y_i}{\sumi w_0(x_i)(1-Z_i)},
\end{equation}
where $w_1(x) =h(x)/{e(x)}, w_0(x) =h(x)/\{1-e(x)\}$ are some type of balancing weights \citep{li2018balancing}, $e(x)=\Pr(Z_i=1|X_i=x)$ is the treatment propensity score, i.e., the conditional probability of treatment given the covariates \citep{Rosenbaum83}, and $h(x)$ is a pre-specified function of the covariates, known as the tilting function \citep{li2019propensity}. In RCT, the true propensity score is known and usually fixed at $e_i(x)=\Pr(Z_i=1)=r$ for all units. In that case, one can show that the weighting estimator $\hat{\tau}^w$ is unbiased for the ATE $\tau$ regardless of the specific choice of $h(x)$ as long as $h(x)$ is a function of the propensity score $e(x)$. Two examples of the balancing weights are (i) inverse probability weights (IPW):  $(w_1, w_0)=(1/e(x), 1/\{1-e(x)\})$, obtained by $h(x)=1$ \citep{williamson2014variance,shen2014inverse,Colantuoni2015}, and (ii) overlap weights (OW): $(w_1, w_0)=(1-e(x), e(x))$, obtained by $h(x)=e(x)(1-e(x))$ \citep{li2018balancing,li2019addressing, zeng2021propensity}. In the context of covariate adjustment, the propensity score $e_i$ in the weighting estimator \eqref{eq:sampleWATE} is replaced by a \emph{working} propensity score $\hat{e}_i=e(X_i;\hat{\theta})$, estimated from a ``working" logistic model \begin{equation}\label{eq:ps_logistic}
e_i={e}(X_i;\theta)=\frac{\exp(\theta_0+X_i^T\theta_1)}{1+\exp(\theta_0+X_i^T\theta_1)},  
\end{equation}
with parameters $\theta=(\theta_0,\theta_1^T)^T$ and $\hat{\theta}$ being the maximum likelihood estimate of $\theta$. For example, the OW estimator for ATE is
\begin{equation}
\label{eq:OWATE}
\hat{\tau}^{\OW}=\frac{\sumi (1-\hat{e}_i)Z_i Y_i}{\sumi (1-\hat{e}_i)Z_i} -
              \frac{\sumi \hat{e}_i(1-Z_i) Y_i}{\sumi \hat{e}_i(1-Z_i)}.
\end{equation} 
OW has a unique \emph{exact balance} property: when the treatment propensity scores are estimated from a logistic model, the weighted average of any predictor in the model is exactly the same between the treatment and control arms \citep{li2018balancing}.

\cite{zeng2021propensity} proved that the weighting estimator $\hat{\tau}^w$ with any balancing weights is asymptotically equivalent to the efficient OLS estimator for continuous outcomes. They further demonstrated via analytical derivations and simulations that the exact balance property of OW improves finite-sample  efficiency in estimating ATE compared to IPW. Therefore, we will focus on OW in the next section. Weighting methods have two practical advantages compared to outcome regression adjustment. First, weighting obviates the need to specify an outcome model and has the same form for all types of outcome. This is particularly desirable in handling rare binary or categorical outcomes, where outcome regression can fail to converge. Second, specification of the treatment propensity score model is usually simpler than that of the outcome model because the former does not include covariate-treatment interactions; this can be beneficial for finite-sample performance when the sample size is small. Third, implementation of weighting methods is more straightforward than outcome regression, which requires centering of all covariates. Moreover, because weighting only involves modeling the baseline data, conceptually it might be more appealing to analysts who would avoid directly modeling the outcome data.

\section{Covariate adjustment with incomplete data} \label{sec:CovAdj_missing}

\subsection{Adjustment with missingness in covariates}\label{sec:missingX}

We introduce additional notation in the presence of incomplete data. We first consider the case of missingness only in covariates. For simplicity, we describe the case with one partially observed covariate $X$, centered based on the observed values, and rest of the covariates are fully observed and centered, denoted as $V$. For each unit $i (i=1,...n)$, let $R^x_{i}$ be the missingness indicator for $X_i$, with $R_i^x = 1$ if the value is observed and 0 if missing. Let $X_i^{\imp}$ denote the imputed value based on some imputation method if the value $X_i$ is missing.  For notational simplicity, below we omitted the fully observed covariate $V$  but stress that one should consider adjustment of imbalanced prognostic variables and stratification factors used in the stratified randomization.  

When there is missingness only in $X$, \cite{zhao2022adjust} showed theoretically that it matters little how to impute the missing $X$ as long as the missingness indicator is included as in the following ANCOVA model for covariate adjustment. Specifically, they showed the minimum specification of regression adjustment to ensure asymptotic efficiency gain is as follows:
\begin{equation} \label{eq:OLS-interaction}
    Y_i \sim Z_i + X_iR^x_i+ R^x_i + Z_i(X_iR^x_i+ R^x_i). 
\end{equation}
This result is proved from a design-based perspective without any assumption on the missing data mechanism of $X$, and the only requirement is that $R^x$ is independent of the treatment assignment. A technical insight is that the interaction term $X_iR^x_i$ is numerically equivalent to (i) $X_i$ when the covariate is observed and (ii) zero when the covariate is missing regardless how the missing value is imputed. Therefore, the term $X_iR^x_i$ effectively imputes the missing values as zero, and we can generically denote $X_i^{\imp}=X_iR^x_i$. In programming, we suggest to use median instead zero as the imputed values because for categorical variables (both lead to identical model fit), the imputed zeros are indistinguishable from the actual zeros, which may cause confusion.     

The theoretical explanation of Zhao and Ding's result lies in Lin's result \citep{lin2013agnostic} that adding more covariates into the OLS adjustment model \eqref{eq:OLS-lin} would always increase the asymptotic efficiency and the fact that the missingness indicators $R^x$ are essentially additional covariates (comparing models \eqref{eq:OLS-lin} and \eqref{eq:OLS-interaction}). We now also provide some empirical intuition for Zhao and Ding's result. Consider a case where there is a covariate that is predictive of a unit's outcome and also his/her propensity to missing data, and that covariate is not fully observed. As a hypothetical example, suppose that older patients are more likely to have a higher outcome value, and also tend to fail to report some covariates, but age is missing for some patients. In this case, the covariate missingness indicator is a strong predictor of the outcome, and thus ensuring its balance would improve the efficiency of the treatment effect estimate. 

Recognizing the importance of the covariate missingness indicators $R^x$, for the propensity score weighting method for covariate adjustment, we suggest to add $R^x$'s into the working propensity score model, as follows:
\begin{equation}\label{eq:ps_logit}
\mbox{logit}\{e(X_i,R_i^x)\} = \theta_0+\theta_1 X_iR^x_i+ \theta_2 R^x_i,
\end{equation} 
and then plug in the estimated propensity score $\hat{e}_i$ to formula \eqref{eq:OWATE} to estimate the treatment effect.

Below we generalize the exact balancing property of overlap weighting to the case with missing covariates. For simplicity in notation, we present the proposition with only one covariate, which is straightforward to be extended to multiple covariates.

\begin{prop}
\label{thm:OW_balance} Assume the working treatment propensity scores are estimated from the logistic regression model \eqref{eq:ps_logit}: $\hat{e}_i= \{1+\exp[-(\hat{\theta}_0 +\hat{\theta}_1 X_iR^x_i+ \hat{\theta}_2 R_i^x)]\}^{-1}$, where $\hat{\theta}$ is the maximum likelihood estimate of the model coefficients. Then the resulting overlap weights lead to exact balance in the means of any predictor, denoted by $W$, in the variable set $(XR^x,R^x)$, between the treatment and control groups. That is,
\begin{equation} \label{eq:exact_balance}
\frac{\sum_i W_{i}Z_i(1-\hat{e}_i)}{\sum_i Z_i(1-\hat{e}_i)}
=\frac{\sum_i W_{i}(1-Z_i)\hat{e}_i}{\sum_i (1-Z_i)\hat{e}_i}.
\end{equation}
\end{prop}
The proof follows the same as that in \cite{li2018balancing} and thus is omitted here. Similar to the case with complete covariates, this exact balance property is the key to improve the finite-sample efficiency over other weighting methods \citep{zeng2021propensity}.

Variance of the covariate-adjusted estimators $\widehat{\tau}^{ols}$ and  $\widehat{\tau}^{w}$ can be estimated via bootstrap, where one resamples the original data with missing values and repeats the analysis detailed above. The bootstrap takes into account the sample variability as well as the uncertainty of imputing the missing data, but it can be computationally intensive. Alternatively, one can use the analytical Eicker-Huber-White robust standard error for the OLS estimator \citep{lin2013agnostic,zhao2022adjust}. This estimator is asymptotically conservative but serves as a convenient approximation to the true standard error. For the weighting estimator, one can use the sandwich variance estimator \citep{lunceford2004stratification,li2019propensity}, both of which are implemented in the R package \textit{PSweight} \citep{zhou2020psweight}.

\subsection{Adjustment with missingness in covariates and outcomes} \label{sec:missingXY} 
We now consider the case of missingness in both covariates and outcome. Let $R^y_{i}$ denote the outcome missingness indicator for unit $i$, with $R_i^y = 1$ if $Y_i$ is observed and 0 if missing. A common method in the presence of missing $Y$ is the \emph{complete-outcome} analysis, which removes all units with missing outcome. Such method implicitly assumes the outcome is \emph{missing completely at random (MCAR)} \citep{rubin1976inference}, that is,  $\Pr(R_i^y=1\mid Y_i,X_i, R_i^x, V_i, Z_i)= \Pr(R_i^y=1)$. This assumption is usually questionable, and can lead to large bias as well as loss of sample size \citep{white2005adjusting}. It should be avoided unless there is a strong reason to believe MCAR holds. A more principled approach is to inversely weigh each observed outcome by its probability of being observed \citep{seaman2013review}. Below we discuss how to proceed with the inverse weighting method in the presence of missing $Y$ under the assumption of outcome \emph{missing at random (MAR)}, that is, the outcome missingness can depend on the fully observed covariates and the missing covariates indicators.  

\begin{assumption} \label{as::MAR-Y1}
    The missing data mechanism of $Y$: $\Pr(R_i^y=1\mid Y_i,X_i, R_i^x, V_i, Z_i)=\Pr(R_i^y=1\mid R_i^x, V_i, Z_i)$.
\end{assumption}
\noindent To proceed with inverse probability weighting, one needs to estimate the probability of $Y_i$ being observed, $p_i=\Pr(R_i^y = 1 \mid R_i^x, Z_i, V_i)$. Similar to the estimation strategy for the treatment propensity score, given Assumption \ref{as::MAR-Y1}, we can postulate a model to estimate the outcome missingness propensity. For example, a simple choice is the following logistic model:
\begin{align} \label{eq:Ry-model-noX}
    \logit \{\Pr(R^y_i=1 \mid R^x_i, V_i, Z_i)\} &=\gamma_0+\gamma_1 R^x_i + \gamma_2 V_i +\gamma_3 Z_i.
\end{align}

We now discuss an alternative MAR outcome missingness mechanism that is weaker than the assumption \ref{as::MAR-Y1}. We assume: \emph{(i)} if the covariate value for unit $i$ is observed, i.e. $R^x_i=1$, then the missingness of $Y_i$ can be dependent on $X_i,Z_i, V_i$: $\Pr(R_i^y=1\mid Y_i,X_i, R_i^x, V_i, Z_i)=\Pr(R_i^y=1\mid X_i, V_i, Z_i)$, and \emph{(ii)} if the covariate value is missing, i.e. $R^x_i=0$, then the missingness of $Y_i$ can be dependent on $Z_i, V_i$: $\Pr(R_i^y=1\mid Y_i,X_i, R_i^x, V_i, Z_i)=\Pr(R_i^y=1\mid V_i, Z_i)$. That is,
\begin{assumption} \label{as::MAR-Y2}
The missing data mechanism of $Y$: $\Pr(R_i^y=1\mid Y_i,X_i, R_i^x, V_i, Z_i)=\Pr(R_i^y=1\mid R_i^x, X_iR^x_{i}, V_i, Z_i)$.
\end{assumption}

Assumption \ref{as::MAR-Y2} is weaker than Assumption \ref{as::MAR-Y1} in the sense that it allows the missingness of $Y$ to also depend on the observed values of the partially observed covariates. This assumption was first made in \cite{rosenbaum1984reducing}. Given Assumption \ref{as::MAR-Y2}, we can postulate the following model to estimate the outcome missingness probability (or propensity score):
\begin{align} \label{eq:Ry-model-X}
    \logit \{\Pr(R^y_i=1 \mid X_i, R^x_i, V_i, Z_i )\} &=\gamma_0+\gamma_1 X_iR^x_i+ \gamma_2 R^x_i + \gamma_3 V_i +\gamma_4 Z_i.
\end{align}
As discussed before, the interaction term $X_iR^x_i$ effectively imputes all the missing values as zero regardless of the actual imputed values.  

The analysts can choose either Assumption \ref{as::MAR-Y1} or \ref{as::MAR-Y2} depending on the specific study, and estimate the outcome missing propensity accordingly. It is straightforward to extend Model \eqref{eq:Ry-model-noX} and \eqref{eq:Ry-model-X} to include multiple partially observed covariates: for each partially observed covariate, one needs to include the corresponding missing indicator $R_i^x$ or/and the interaction term $X_iR_i^x$ as predictors. Note that the role of the outcome missingness propensity ($p_i$) is different from that of the ``working'' treatment propensity ($e_i$). Specifically, the treatment propensity score is used to adjust for chance imbalance in baseline characteristics to improve efficiency and it is never misspecified because of the randomization. On the other hand, the outcome propensity score is used to adjust for potential bias due to informative missingness in outcomes and its validity depends on the correct specification of the outcome missingness process. 

Once we estimate the outcome missingness and the treatment propensity (denoted by $\hat{p}_i$ and $\hat{e}_i$, respectively), there are two approaches for covariate adjustment. The first approach is to conduct a weighted least square (WLS) by weighting the ANCOVA \eqref{eq:OLS-interaction} on the observed outcomes with the weights $\widehat{p}_i$, and the estimated coefficient of $Z$ is the treatment effect $\widehat{\tau}^{wls}$. Below we instead focus on the weighting approach following the discussion in Section \ref{sec:CovAdj}, partly due to the coherency of a full weighting approach. Specifically, we propose a full weighting approach that combines overlap weighting for covariate adjustment and inverse probability weighting for adjusting for the missing outcomes, as follows. 
\begin{enumerate}
    \item[S1.] Impute the missing covariate values by zeros or median of the observed values of that covariate.
    \item[S2.] Estimate the outcome missingness propensity for each unit $\widehat{p}_i$, e.g. via Model \eqref{eq:Ry-model-noX} or  \eqref{eq:Ry-model-X}.
    \item[S3.] Estimate the working treatment propensity score for each unit $\widehat{e}_i$, via the logistic model \eqref{eq:ps_logit}.
    \item[S4.] Estimate the covariate-adjusted treatment effect by an inverse weighted Haj\'ek estimator based on units with observed outcomes
    \begin{equation}
\label{eq:Hajek}
\widehat{\tau}^{ow-ipw}=\frac{\sumi w_1(X_i) R^y_iZ_i Y_i}{\sumi w_1(X_i)R^y_iZ_i} -
              \frac{\sumi w_0(X_i)R^y_i(1-Z_i) Y_i}{\sumi w_0(X_i)R^y_i(1-Z_i)},
\end{equation}
where the weight $w$ is the product of the overlap weight based on $\widehat{e}_i$ and the inverse of the missingness propensities $\widehat{p}_i$: $w_1(X_i)=(1-\widehat{e}_i)/\widehat{p}_i$ and $w_0(X_i)=\widehat{e}_i/\widehat{p}_i$.  
\end{enumerate}

The variance of the covariate-adjusted estimators $\widehat{\tau}^{ow-ipw}$ can be estimated via bootstrap, where one resamples the original data with missing values and repeats the analysis detailed above. The bootstrap takes into account the sample variability, the uncertainty of imputing the missing data, as well as the uncertainty in estimating both the treatment and outcome missingness propensity scores.

\section{Simulations} \label{sec:simulations}
We conduct extensive simulations to investigate the finite-sample performance of the proposed estimator with a variety of alternatives. In particular, we examine these questions: (i) Is the choice of imputation methods important for efficiency? Does misspecified imputation models decrease efficiency?  (ii) Does the asymptotic equivalence between outcome regression and propensity score weighting extend to the case with missing covariates and translate into finite-sample similarities? (iii) Is it important to account for the missingness in the outcome data at the same time? We first conduct simulation studies with missing covariates only. We then extend the simulation to incorporate both missing covariates and outcomes.

\subsection{Simulations with missing covariates and fully observed outcomes}
\label{sim:mis-x-full-y}
\subsubsection{Continuous outcome}
\label{sim:mis-x-cont-y}
This section focuses on the setting with fully observed continuous outcomes. We consider a setting with $p=3$ baseline covariates, $\mathbf{X}_i =(X_{i1},X_{i2},X_{i 3})^T$, among which $X_1$ is highly predictive of the outcome but subject to missingness, while $X_2$, $X_3$ are less predictive and completely observed. In practice, this could correspond to the scenario where $X_1$ is the baseline measurement of the outcome, and $X_2$ and $X_3$ are some other less important covariates. Covariates $X_{i1},X_{i2}$ are generated from a bivariate normal distribution with  mean $(0,0)^T$, variance $(1,1)^T$, and correlation $0.3$; $X_{i3}$ is simulated from a centered Bernoulli distribution, i.e. $\mbox{Bern(0.5)}-0.5$. Let $R_{i}^{x_1}$ denote the missing indicator of $X_{i1}$ and $R_{i}^{x_1}=1$ if $X_{i1}$ is observed and $0$ otherwise. 
We focus on simple randomization of sample size $N=100$ or $500$ with a balanced design $(r=1/2)$.  The potential outcomes are generated from the following linear model: for $z=0,1$
\begin{align}
    &Y_i(z)\sim \mathcal{N}(\beta_0 + \alpha z+\mathbf{X}_i^T\boldsymbol{\beta}_1+ z \mathbf{X}_i^T\boldsymbol{\beta}_2, \sigma_y^2)
\end{align}
The observed outcome is set to be $Y_i = Y_i(Z_i) = Z_iY_i(1) + (1-Z_i)Y_i(0)$. Since all covariates have mean 0, the coefficient $\alpha$ corresponds to the true ATE (i.e., $\tau$) on the additive scale, and we fix $\alpha=\tau=0$. We set the intercept $\beta_0=0.8$ and the covariate main effects  $\boldsymbol\beta_1 = ( 3, 0.3, 0.42)$ and the treatment-covariate interaction effects $\boldsymbol\beta_2=(0.75, 0.53, 0.38)$. The residual variance $\sigma_y^2$ is set to $1$. \\ 
\indent We generate data based on the following three types of missing data mechanisms: 

\begin{enumerate}
    \item Missing completely at random (MCAR): $R_{i}^{x_1}\overset{iid}{\sim} \mbox{Bern}(q)$
    \item Missing at random (MAR):
    $\mbox{logit}\{
    \Pr(R_{i}^{x_1}=1)\} = \gamma_0 + \gamma_1X_{i2} +\gamma_2X_{i3}$
    \item Missing not at random (MNAR):  $\mbox{logit}\{\Pr(R_{i}^{x_1}=1)\} = \zeta_0 + \zeta_1X_{i1}$
\end{enumerate}
We fix $\gamma_1= \gamma_2 = \zeta_1 = -1$, and for each missing mechanism, we simulate data with approximately $10\%$ or $30\%$ of missingness in $X_{i1}$ through varying the values of $q,\gamma_0,\zeta_0$ respectively.  \\
\indent To summarize, our simulation scenarios consist of $2\times 2\times 3 = 12$ combinations of the total sample size (2 levels), the covariates missing proportion (2 levels) and the missing data mechanism (3 levels), and for each scenario, we conduct covariate adjustment using both OW and ANCOVA. We compare three types of imputation methods: (1) mean imputation; (2) correct model imputation where we impute with the predicted values from the linear model $X_{1} \sim X_2$ fitted to those with $X_1$ observed; (3) wrong model imputation where we impute with the predicted values from the linear model $X_{1} \sim X_2^2 + X_3$. For each type of imputation methods, covariate adjustment is performed both with and without the missing indicators. That is, for ANCOVA, we compare the model $Y_i \sim Z_i + X_{i1}^{imp} +X_{i2} + X_{i3} + Z_i(X_{i1}^{imp} + X_{i2} + X_{i3})$ with $Y_i \sim Z_i + X_{i1}^{imp} +X_{i2} + X_{i3} + R_{i}^{x_1} + Z_i(X_{i1}^{imp} + X_{i2} + X_{i3} +R_{i}^{x_1})$. And for OW, the treatment propensity score model without missing indicator is $\mbox{logit}\{Pr(Z=1)\} \sim X_{i1}^{imp} + X_{i2} + X_{i3}$, and with missing indicator it becomes $\mbox{logit}\{Pr(Z_i=1)\} \sim X_{i1}^{imp} + X_{i2} + X_{i3} + R_{i}^{x_1}$. Additionally, we include two types of complete data analysis (complete-unit and complete-covariate), as well as an unattainable ideal case where ANCOVA/OW is applied to the full data with no missingness. The resulting estimators of $\tau$ are evaluated in terms of their empirical bias and efficiency. The relative efficiency is defined as the ratio between the Monte Carlo variance of the unadjusted estimator to that of the adjusted estimator. The higher the relative efficiency, the more efficient that estimator is compared to the unadjusted estimator. 

Table \ref{con100} summarizes the simulation results for sample size $N=100$. All approaches produce nearly unbiased estimates of the ATE except for the complete-unit approach under the MAR and MNAR settings, where the magnitude of bias increases as the proportion of missingness increases from 10\% to 30\%.  Overall, performing covariate adjustment using either ANCOVA or OW improves efficiency over the unadjusted estimators, and adjustment with imputed $X_1$ results in further efficiency gain over adjustment with complete covariates only. On the other hand, comparisons among different imputation methods render minimal differences, which echoes the theoretical results by \cite{zhao2022adjust}. Inclusion of the missing indicator in the ANCOVA or the propensity score model does not appear to impact efficiency much under MCAR and MAR, as seen in \cite{kayembe2022imputation}, but can substantially improve efficiency under the MNAR setting. Similar trends are observed for sample size $N=500$ and the results are provided in Table \ref{con500}.
\begin{landscape}
\begin{table}[H]
    \centering
    \begin{tabular}{lcccccccccccc}
    \toprule
    & \multicolumn{12}{c}{Covariates Missingness Mechanism} \\ [1pt]
    \cline{2-13} \\ [-8pt]
     & \multicolumn{4}{c}{MCAR} & \multicolumn{4}{c}{MAR} & \multicolumn{4}{c}{MNAR} \\ 
         &  \multicolumn{2}{c}{Bias} & \multicolumn{2}{c}{RE} & \multicolumn{2}{c}{Bias} & \multicolumn{2}{c}{RE} & \multicolumn{2}{c}{Bias} & \multicolumn{2}{c}{RE} \\
    Method     & ANCOVA & OW & ANCOVA & OW & ANCOVA & OW & ANCOVA & OW& ANCOVA & OW& ANCOVA & OW\\
    \midrule
     \multicolumn{13}{l}{$N=100$, $30\%$ missing} \\
    Full data &     .00    &.00  &   11.22   &  11.04 &.00  &.00    &     11.22     &11.04      &.00  &.00  &       11.22   &  11.04  \\
    Complete covariate &    .02  &  .02  &        1.20    &  1.20   &     .02   &.02    &      1.20    &  1.20        &.02    &.02        &  1.20      &1.20 \\
    Complete unit &   .00    &.00     &     7.78   &   7.66  &      .20    &.20      &    7.77   &   7.69    &.23    &.23      &    7.72   &   7.59
\\
    Mean imputation w/o MSI &  .01    &.01      &    2.92    &  2.92        &.01    &.01        &  2.98   &   2.97        &.00    &.00     &     2.44   &   2.45 \\
    Mean imputation w/ MSI &   .01    &.01   &       2.91     & 2.92        &.01    &.01      &    2.99  &    3.00        &.00    &.00      &    3.42 &     3.45\\
    Correct model imputation w/o MSI & .01    &.01    &      3.05   &   3.04        &.01    &.01     &     3.08  &    3.08       &.00    &.01      &    2.60   &   2.61\\ 
    Correct model imputation w/ MSI &    .01    &.01  &        3.05   &   3.05        &.01   &.00      &    3.07   &   3.08    &.00    &.00     &     3.58    &  3.60 \\
    Wrong model imputation w/o MSI &  .01   &.01    &      2.89  &    2.86        &.01    &.01      &    2.88  &    2.85        &.00   & .01     &     2.86  &    2.42 \\
    Wrong model imputation w/ MSI &     .01    &.01     &     2.85      &2.86        &.01    &.01    &      2.85  &    2.87        &.00    &.00       &   3.37   &   3.40 \\ 
\\
     \multicolumn{13}{l}{$N=100$, $10\%$ missing} \\
    Full data &  .00    &.00    &     11.22   & 11.04      &  .00    &.00        & 11.22    & 11.04        &.00    &.00        & 11.22   &  11.04\\
    Complete covariate &.02   &.02       &   1.20    &  1.20        &.02   &.02      &    1.20    &  1.20     &  .02    &.02         & 1.20     & 1.20\\
    Complete unit &.00    &.00    &     10.12   &   9.93   &     .07   &.07      &   10.17    &  9.95    &.08    &.08        & 10.32  &   10.13\\
    Mean imputation w/o MSI &.00   &.00    &     5.70    &  5.64       &.00  &.00       &   5.61     & 5.54        &.00    &.00     &     4.42   &   4.40\\
    Mean imputation w/ MSI &.00  &.00    &      5.44     & 5.83     &.00   &.00      &    5.52    &  5.86 &.00  &.00 &        5.90    &  6.31\\
    Correct model imputation w/o MSI &.00   &.00    &      5.93    &  5.87        &.00    &.00       &   5.94    &  5.86       &.00    &.00         & 4.72  &    4.69
 \\
    Correct model imputation w/ MSI &.00    &.00     &     5.68     & 6.04       &.00   &.00       &   5.73    &  6.04       &.00   &.00     &     6.12   &   6.52\\
    Wrong model imputation w/o MSI &.00   &.00   &       5.68     & 5.58      &.00    &.01    &      5.47   &   5.34        &.00    &.00       &   4.66   &   4.36\\
    Wrong model imputation w/ MSI &.00    &.00     &     5.38  &    5.76        &.00    &.00      &    5.32   &   5.68  &      .00    &.00      &    5.83  &    6.24\\
    \bottomrule
    \end{tabular}
    \caption{Simulation results for continuous outcomes with sample size 100. RE stands for relative efficiency compared to the unadjusted estimator; MSI stands for missing indicators. The results are based on 5000 iterations. }
    \label{con100}
\end{table}
\end{landscape}

\begin{landscape}
\begin{center}
\end{center}
\begin{table}[H]
    \centering
    \begin{tabular}{lcccccccccccc}
    \toprule
     & \multicolumn{4}{c}{MCAR} & \multicolumn{4}{c}{MAR} & \multicolumn{4}{c}{MNAR} \\ 
      &  \multicolumn{2}{c}{Bias} & \multicolumn{2}{c}{RE} & \multicolumn{2}{c}{Bias} & \multicolumn{2}{c}{RE} & \multicolumn{2}{c}{Bias} & \multicolumn{2}{c}{RE} \\
     Method     & ANCOVA & OW & ANCOVA & OW & ANCOVA & OW & ANCOVA & OW& ANCOVA & OW& ANCOVA & OW\\
     \midrule
      \multicolumn{13}{l}{$N=500$, $30\%$ missing} \\
     Full data     &.00    &.00    &     11.00    & 10.95       &.00    &.00        & 11.00     &10.95       &.00    &.00     &    11.00     &10.95   \\
     Complete covariate   &.01    &.01         & 1.20    &  1.20        &.01    &.01      &    1.20 &     1.20        &.01    &.01         & 1.20     & 1.20  \\
     Complete unit  &.00    &.00      &    7.88  &    7.85        &.21    &.21        & 7.97   &   7.92        &.23    &.23 &         7.77    &  7.74\\
     Mean imputation w/o MSI &.00    &.00      &    2.92  &    2.92        &.00    &.00     &     2.96   &   2.95        &.01    &.01         & 2.50     & 2.49 \\
     Mean imputation w/ MSI    &.00    &.00         & 2.91    &  2.92        &.00    &.00        &  3.00   &   3.00       &.00    &.00        &  3.41   &   3.41\\
     Correct model imputation w/o MSI  &.00    &.00      &    3.06   &   3.06    &.00   &.00       &   3.12  &    3.12       &.01  &.01  &        2.60   &   2.60\\ 
     Correct model imputation w/ MSI &.00    &.00        &  3.06     & 3.06        &.00    &.00       &   3.13    &  3.13        &.00    &.00        &  3.49    &  3.49\\
     Wrong model imputation w/o MSI   &.00    &.00       &   2.90    &  2.90        0&.00    &.00          &2.89   &   2.86       &.00    &.01        &  2.90   &   2.48 \\
     Wrong model imputation w/ MSI  &.00    &.00     &     2.89    &  2.90     &   .00    &.00    &      2.90 &     2.90        &.00    &.00      &    3.39   &   3.39\\ 
 \\
      \multicolumn{13}{l}{$N=500$, $10\%$ missing} \\
     Full data  &.00     &.00         & 11.00     & 10.95      &.00     &.00         & 11.00    &  10.95       &.00     &.00      &    11.00      &10.95\\
     Complete covariate   &.01     &.01       &    1.20    &   1.20         &.01     &.01          & 1.20       &1.20         &.01   &.01          & 1.20      & 1.20\\
     Complete unit  &.00     &.00        &   9.86     &  9.81     &.07     &.07         & 10.01      & 9.96       &.08   &.08      &    10.05     & 10.02\\
     Mean imputation w/o MSI  &.00     &.00        &   5.73    &   5.71        &.00     &.00       &    5.65      & 5.63        &.00     &.00           &4.50       &4.49\\
     Mean imputation w/ MSI   &.00     &.00       &    5.66     &  5.74         &.00    &.00        &   5.76      & 5.82         &.00     &.00       &    6.21    &   6.28\\
     Correct model imputation w/o MSI  &.00    &.00       &    5.98    &  5.96       &.00     &.00        &   6.00   &    5.98  &.00    &.00        &   4.82     &  4.81\\
     Correct model imputation w/ MSI   &.00     &.00    &       5.91    &   5.98      &.00     &.00       &    5.98   &    6.04     &.00     &.00        &   6.44    &   6.51\\
     Wrong model imputation w/o MSI  &.00     &.00          & 5.71     &  5.70     &.00     &.00     &      5.58   &    5.53       &.00     &.00      &     4.73    &   4.47\\
     Wrong model imputation w/ MSI  &.00     &.00     &      5.64      & 5.72        &.00     &.00         &  5.67     &  5.73        &.00     &.00      &     6.18    &   6.26 \\ 
     \bottomrule
     \end{tabular}
     \caption{Simulation results for continuous outcomes with sample size 500. RE stands for relative efficiency compared to the unadjusted estimator; MSI stands for missing indicators. The results are based on 5000 iterations.}
     \label{con500}
\end{table}
\end{landscape}

\subsubsection{Binary outcome}
\label{sim:mis-x-bin-y}
We perform another set of simulations under the setting with fully observed binary outcomes. We consider a balanced randomization design ($r=0.5$) and use the same covariate generating processes and missingness processes as in Section \ref{sim:mis-x-cont-y}. The potential outcomes are generated from the following logistic model: for $z=0,1,$
\begin{align}
    \mbox{logit}\{\mbox{Pr}(Y_i(z)=1)\} = \beta_0 + \alpha z +\mathbf{X}_i^T\boldsymbol{\beta}_1+ z \mathbf{X}_i^T\boldsymbol{\beta}_2
\end{align}
We fixed $\beta_0=\alpha=0$ and $\boldsymbol{\beta}_1=(4,1,1)^T, \boldsymbol{\beta}_2=(-3.5,0.3,0.3)^T$. Same as before, our target estimand $\tau$ is the ATE on the additive scale, namely the marginal risk difference between the treatment and control arms. Under the current parameter setup, we have $\tau=0$. The simulations for binary outcomes follow the same $2\times 2\times 3$ factorial design as for the continous outcomes. \\ 
\indent  Non-collapsibility of odds ratio is a well-known issue for binary outcomes \citep{greenland1999confounding}. If one applies ANCOVA-type logistic regression model to estimate the marginal risk difference, one needs to first estimate the group means by standardization using the coefficients obtained from the fitted model. To avoid such complication, we conduct ANCOVA-type covariate adjustment through the standard linear regression, which allows us to directly interpret the coefficient of the treatment indicator as the marginal risk difference. In contrast, implementation of OW is more straightforward and the ATE estimator takes the form in eq (4) for both continuous and binary outcomes. Similar to Section \ref{sim:mis-x-cont-y}, we compare the same types of imputation methods (i.e., mean imputation, correctly specified, and incorrectly specified model-based imputation), two types of complete data analysis (i.e., complete-unit and complete-covariate), and the unattainable ideal case (i.e., application of ANCOVA/OW to the full data with no missingness). We again consider both bias and the relative efficiency for each estimator to the unadjusted estimator.

The results for binary outcomes are presented in Table \ref{bin100} for $N= 100$ and in Table \ref{bin500} for $n = 500$. In general, results from binary outcomes are similar to those from the continuous outcome settings. The ANCOVA and OW estimators perform similarly for all settings with negligible differences. The complete-unit analyses can be biased when covariates are MAR or MNAR. The complete-covariate analyses are seen with reduced efficiency in these settings due to the omission of the partially-observed covariate $\boldsymbol{X}_1$ that is predictive of the outcome. Incorporating $\boldsymbol{X}_1$ in estimating the ATE in general improves efficiency, even when the imputation model is misspecified. When covariates are MNAR, including missingness indicator in the adjustment model can lead to further efficiency improvement. 

\begin{landscape}
\begin{table}[H]
    \centering
    \begin{tabular}{lcccccccccccc}
    \toprule
    & \multicolumn{12}{c}{Covariates Missingness Mechanism} \\ [1pt]
    \cline{2-13} \\ [-8pt]
     & \multicolumn{4}{c}{MCAR} & \multicolumn{4}{c}{MAR} & \multicolumn{4}{c}{MNAR} \\ 
         &  \multicolumn{2}{c}{Bias} & \multicolumn{2}{c}{RE} & \multicolumn{2}{c}{Bias} & \multicolumn{2}{c}{RE} & \multicolumn{2}{c}{Bias} & \multicolumn{2}{c}{RE} \\
    Method     & ANCOVA & OW & ANCOVA & OW & ANCOVA & OW & ANCOVA & OW& ANCOVA & OW& ANCOVA & OW\\
    \midrule
     \multicolumn{13}{l}{$N=100$, $30\%$ missing} \\
    Full data   & .00     & .00     & 1.53  & 1.52  & .00     & .00     & 1.53  & 1.52  & .00     & .00     & 1.53  & 1.52 \\
    Complete covariate    & .00     & .00     & 1.22  & 1.22  & .00     & .00     & 1.22  & 1.22  & .00     & .00     & 1.22  & 1.22 \\
    Complete unit    & .00     & .00     & 1.02  & 1.01  & .02  & .03  & 1.02  & 1.01  & .06  & .06  & 1.03  & 1.03 \\
    Mean imputation w/o MSI   & .00     & .00     & 1.40   & 1.39  & .00     & .00     & 1.41  & 1.40   & .00     & .00     & 1.38  & 1.37 \\
    Mean imputation w/ MSI   & .00     & .00     & 1.39  & 1.38  & .00     & .00     & 1.40   & 1.40   & .00     & .00     & 1.43  & 1.42 \\
    Correct model imputation w/o MSI   & .00     & .00     & 1.41  & 1.40   & .00     & .00     & 1.42  & 1.41  & .00     & .00     & 1.39  & 1.38 \\
    Correct model imputation w/ MSI   & .00     & .00     & 1.40  & 1.39  & .00     & .00     & 1.40   & 1.40   & .00     & .00     & 1.43  & 1.42 \\
    Wrong model imputation w/o MSI   & .00    & .00     & 1.39  & 1.38  & .00     & .00     & 1.41  & 1.40   & .00     & .00     & 1.38  & 1.37 \\
    Wrong model imputation w MSI   & .00     & .00     & 1.38  & 1.38  & .00     & .00     & 1.39  & 1.39  & .00     & .00     & 1.42  & 1.41 \\
    \\
     \multicolumn{13}{l}{$N=100$, $10\%$ missing} \\
    Full data    & .00     & .00     & 1.53  & 1.52  & .00     & .00     & 1.53  & 1.52  & .00     & .00     & 1.53  & 1.52 \\
    Complete covariate    & .00     & .00     & 1.22  & 1.22  & .00     & .00     & 1.22  & 1.22  & .00     & .00     & 1.22  & 1.22 \\
    Complete unit    & .00     & .00     & 1.36  & 1.35  & .01  & .01  & 1.35  & 1.34  & .02  & .02  & 1.35  & 1.34 \\
    Mean imputation w/o MSI   & .00     & .00     & 1.49  & 1.48  & .00  & .00   & 1.49  & 1.48  & .00 & .00 & 1.46  & 1.45 \\
    Mean imputation w/ MSI   & .00 & .00 & 1.47  & 1.47  & .00 & .00 & 1.48  & 1.48  & .00 & .00 & 1.48  & 1.48 \\
    Correct model imputation w/o MSI   & .00 & .00     & 1.49  & 1.48  & .00 & .00 & 1.50  & 1.48  & .00 & .00     & 1.47  & 1.46 \\
    Correct model imputation w/ MSI   & .00     & .00     & 1.47  & 1.47  & .00     & .00     & 1.48  & 1.48  & .00     & .00     & 1.48  & 1.48 \\
    Wrong model imputation w/o MSI   & .00     & .00     & 1.49  & 1.47  & .00     & .00     & 1.49  & 1.48  & .00     & .00     & 1.46  & 1.45 \\
    Wrong model imputation w/ MSI  & .00     & .00     & 1.46  & 1.46  & .00     & .00     & 1.47  & 1.47  & .00     & .00     & 1.48  & 1.48 \\
    \bottomrule
    \end{tabular}
    \caption{Simulation results for binary outcomes with sample size 100. RE stands for relative efficiency compared to the unadjusted estimator; MSI stands for missing indicators. The results are based on 5000 iterations.}
    \label{bin100}
\end{table}
\end{landscape}

\begin{landscape}
\begin{center}
\end{center}
\begin{table}[H]
    \centering
    \begin{tabular}{lcccccccccccc}
    \toprule
     & \multicolumn{4}{c}{MCAR} & \multicolumn{4}{c}{MAR} & \multicolumn{4}{c}{MNAR} \\ 
      &  \multicolumn{2}{c}{Bias} & \multicolumn{2}{c}{RE} & \multicolumn{2}{c}{Bias} & \multicolumn{2}{c}{RE} & \multicolumn{2}{c}{Bias} & \multicolumn{2}{c}{RE} \\
     Method     & ANCOVA & OW & ANCOVA & OW & ANCOVA & OW & ANCOVA & OW& ANCOVA & OW& ANCOVA & OW\\
     \midrule
      \multicolumn{13}{l}{$N=500$, $30\%$ missing} \\
    Full data    & .00     & .00     & 1.51  & 1.51  & .00     & .00     & 1.51  & 1.51  & .00     & .00     & 1.51  & 1.51 \\
    Complete covariate   & .00     & .00     & 1.23  & 1.23  & .00     & .00     & 1.23  & 1.23  & .00     & .00     & 1.23  & 1.23 \\
    Complete unit    & .00     & .00     & 1.06  & 1.05  & .02  & .02  & 1.04  & 1.04  & .06  & .06  & 1.05  & 1.05 \\
    Mean imputation w/o MSI   & .00     & .00     & 1.42  & 1.42  & .00     & .00     & 1.42  & 1.42  & .00     & .00     & 1.41  & 1.41 \\
    Mean imputation w/ MSI   & .00     & .00     & 1.42  & 1.42  & .00     & .00     & 1.42  & 1.42  & .00     & .00     & 1.45  & 1.45 \\
    Correct model imputation w/o MSI   & .00     & .00     & 1.42  & 1.42  & .00     & .00     & 1.42  & 1.42  & .00     & .00     & 1.41  & 1.41 \\
    Correct model imputation w/ MSI  & .00     & .00     & 1.42  & 1.42  & .00     & .00     & 1.42  & 1.42  & .00     & .00     & 1.45  & 1.45 \\
    Wrong model imputation w/o MSI   & .00     & .00     & 1.42  & 1.42  & .00     & .00     & 1.42  & 1.42  & .00     & .00     & 1.41  & 1.41 \\
    Wrong model imputation w/MSI   & .00     & .00     & 1.42  & 1.42  & .00     & .00     & 1.42  & 1.42  & .00     & .00     & 1.45  & 1.45 \\
 \\
      \multicolumn{13}{l}{$N=500$, $10\%$ missing} \\
    Full data    & .00     & .00     & 1.51  & 1.51  & .00     & .00     & 1.51  & 1.51  & .00     & .00     & 1.51  & 1.51 \\
    Complete covariate    & .00     & .00     & 1.23  & 1.23  & .00     & .00     & 1.23  & 1.23  & .00     & .00     & 1.23  & 1.23 \\
    Complete unit    & .00     & .00     & 1.36  & 1.36  & .01  & .01  & 1.36  & 1.35  & .02  & .02  & 1.36  & 1.35 \\
    Mean imputation w/o MSI  & .00     & .00     & 1.47  & 1.47  & .00     & .00     & 1.49  & 1.48  & .00     & .00     & 1.48  & 1.48 \\
    Mean imputation w/ MSI  & .00     & .00     & 1.47  & 1.47  & .00     & .00     & 1.49  & 1.48  & .00     & .00     & 1.50   & 1.50 \\
    Correct model imputation w/o MSI & .00     & .00     & 1.47  & 1.47  & .00     & .00     & 1.48  & 1.48  & .00     & .00     & 1.49  & 1.49 \\
    Correct model imputation w/ MSI   & .00     & .00     & 1.47  & 1.47  & .00     & .00     & 1.48  & 1.48  & .00     & .00     & 1.50  & 1.50 \\
    Wrong model imputation w/o MSI   & .00     & .00     & 1.47  & 1.47  & .00     & .00     & 1.49  & 1.48  & .00     & .00     & 1.48  & 1.48 \\
    Wrong model imputation w/ MSI   & .00     & .00     & 1.47  & 1.47  & .00     & .00     & 1.49  & 1.48  & .00     & .00     & 1.50   & 1.50 \\ 
     \bottomrule
     \end{tabular}
     \caption{Simulation results for binary outcomes with sample size 500. RE stands for relative efficiency compared to the unadjusted estimator; MSI stands for missing indicators. The results are based on 5000 iterations.}
     \label{bin500}
\end{table}
\end{landscape}

\subsection{Simulations with missing covariates and missing outcomes}
\label{sim:mis-x-mis-cont-y}
\begin{align} \label{eq:Ry-model}
logit(\Pr(R_i^y = 1|R_i^{x_1}, Z_i, X_{i2}, X_{i3})) = \theta_0 + \theta_{Z} Z_i + \theta_{Z R^{x_1^*}} Z_i R_i^{x_1^*} + \sum_{j=2}^{3}\theta_{Z X_{j}} Z_i X_{ij}, 
\end{align}
where $R^{x_1^*} = R^{x_1} - \overline{R^{x_1}}$ is the centered covariate missingness indicator. We fixed $\boldsymbol{\theta} =(\theta_0, \theta_Z, \theta_{ZR^{x_1^*}}, \theta_{ZX_2}, \theta_{ZX_3}) = (0.8, 1, 1, 1, 1)$, yielding around $25\%$ of missing outcomes. To demonstrate the importance of correcting potential bias due to informative missing outcomes, we first ignore covariate adjustment and compare the complete-outcome analysis (i.e., difference-in-mean estimator based on units with complete outcomes) with the standard IPW method using the correctly specified propensity score model for outcome missingness. We further investigate different strategies in dealing with missing covariates based on our proposed weighting estimator (\ref{eq:Hajek}). In particular, we correctly specify the propensity score model for outcome missingness and consider treatment propensity score models under the complete-covariate analysis (i.e., ignoring $X_1$), the unattainable ideal case (i.e., using full covariates data), and the same covariates imputation methods from Section \ref{sim:mis-x-full-y}. The performance of each estimator is then evaluated by its empirical bias and efficiency compared to the IPW estimator without covariate adjustment. 

Table \ref{cont100-mis-y} shows the simulation results for $N=100$. As anticipated, the complete-outcome analysis without the IPW adjustment ignores informative missing outcomes, which results in bias across all settings. However, the bias disappears with the IPW adjustment, irrespective of the methods used to adjust for covariates. When the treatment propensity score model fails to account for $X_1$, the efficiency gain is reduced compared to estimators that include imputed $X_1$ in the model, even when $X_1$ is imputed under the misspecified model. When covariates are MNAR, including covariate missingness indicators in the adjustment model always leads to further efficiency gain, consistent with the simulation results in Section \ref{sim:mis-x-cont-y} and \ref{sim:mis-x-bin-y}. Similar trends are observed for sample size $N = 500$, and the results are presented in Table \ref{cont500-mis-y}.

\begin{table}[H]
    \centering
    \resizebox{\columnwidth}{!}{
    \begin{tabular}{lcccccc}
    \toprule
     & \multicolumn{6}{c}{Covariates Missingness Mechanism} \\ [1pt]
     \cline{2-7} \\ [-8pt]
     & \multicolumn{2}{c}{MCAR} & \multicolumn{2}{c}{MAR} & \multicolumn{2}{c}{MNAR} \\ 
         &  Bias & RE  &  Bias & RE &  Bias & RE \\
        \midrule
     \multicolumn{7}{l}{\underline{$N=100$, $30\%$ missing X}}\\
    Complete outcome w/o cov. adj.  & .34  &     1.08  & .29  & 1.07  & .23  & 1.06 \\
    IPW w/o cov. adj. & .03 & 1.00 & .02 & 1.00 & .01 & 1.00 \\
    IPW \& Full X  & .01 & 2.62 & .01 & 2.81 & .00 & 2.74     \\
    IPW \& Complete covariate & .02 & 1.10 & .02 & 1.11 & .01 & 1.12 \\
    IPW \& Mean imputation w/o MSI  & .02 & 1.77 & .02 & 1.91 & .01 & 1.68  \\ 
    IPW \& Mean imputation w/ MSI  &  .03 & 1.77 & .02 & 1.91 & .00 & 1.92  \\ 
    IPW \& Correct model imputation w/o MSI  & .02 & 1.81 & .02 & 1.93 & .01 & 1.72  \\ 
    IPW \& Correct model imputation w/ MSI  & .02 & 1.81 & .02 & 1.93 & .00 & 1.94  \\
    IPW \& Wrong model imputation w/o MSI   & .02 & 1.77 & .02 & 1.86 & .00 & 1.67   \\
    IPW \& Wrong model imputation w/ MSI   & .03 & 1.77 & .02 & 1.87 & .00 & 1.91 \\ 

    \multicolumn{7}{l}{\underline{$N=100$, $10\%$ missing X}} \\
    Complete outcome w/o cov. adj. & .34    & 1.10  & .33  & 1.07  & .30  & 1.07  \\
    IPW w/o cov. adj. & .03 & 1.00 & .03 & 1.00 & 0.02 & 1.00 \\
    IPW \& Full X  & .02  & 2.69 & .01 & 2.76 & .00 & 2.68  \\
    IPW \& Complete covariate & .03 & 1.11 & .03 & 1.12 & .02 & 1.12 \\
    IPW \& Mean imputation w/o MSI  & .02 & 2.33 & .01 & 2.35 & .01 & 2.16  \\ 
    IPW \& Mean imputation w/ MSI  & .02 &     2.36 & .01 & 2.37 & .00 & 2.41 \\ 
    IPW \& Correct model imputation w/o MSI  & .02 & 2.36 & .01 & 2.39 & .01 & 2.20 \\ 
    IPW \& Correct model imputation w/ MSI  & .02  & 2.39 & .01 & 2.39 & .00 & 2.42  \\
    IPW \& Wrong model imputation w/o MSI   &  .02 & 2.32 & .01 & 2.34 & .01 & 2.15  \\
    IPW \& Wrong model imputation w/ MSI   & .02 & 2.35 & .01 & 2.36 & .00 & 2.40 \\ 
    \bottomrule
    \end{tabular}
    }
    \caption{Simulation results with missing covariates and missing continuous outcomes (N=100, 5000 iterations). RE stands for relative efficiency compared to the IPW estimator without covariate adjustment; MSI stands for missing indicators.}
    \label{cont100-mis-y}
\end{table}

\begin{table}[H]
    \centering
    \resizebox{\columnwidth}{!}{
    \begin{tabular}{lcccccc}
    \toprule
     & \multicolumn{6}{c}{Covariates Missingness Mechanism} \\ [1pt]
     \cline{2-7} \\ [-8pt]
     & \multicolumn{2}{c}{MCAR} & \multicolumn{2}{c}{MAR} & \multicolumn{2}{c}{MNAR} \\ 
         &  Bias & RE  &  Bias & RE &  Bias & RE \\
        \midrule
     \multicolumn{7}{l}{\underline{$N=500$, $30\%$ missing X}}\\
    Complete outcome w/o cov. adj.  & .35 &     1.08 & .29 & 1.03 & .23 & 1.03 \\
    IPW w/o cov. adj. & .00 & 1.00 & .00 & 1.00 & .00 & 1.00   \\
    IPW \& Full X  & .00 & 2.84 & .00 & 3.20  & .00 & 3.10  \\
    IPW \& Complete covariate     & .00 & 1.13  & .00 & 1.15 & .00 & 1.14 \\
    IPW \& Mean imputation w/o MSI  & .00 & 1.95 & .00 & 2.01 & .00 & 1.82 \\ 
    IPW \& Mean imputation w/ MSI  &  .00 & 1.96 & .00 & 2.02 & .00 & 2.14 \\ 
    IPW \& Correct model imputation w/o MSI  & .00 & 2.00 & .00 & 2.06 & .00 & 1.87 \\ 
    IPW \& Correct model imputation w/ MSI  & .00 & 2.00 & .00 & 2.06 & .00 & 2.17  \\
    IPW \& Wrong model imputation w/o MSI   & .00 & 1.95 & .00 & 1.97 & .00 & 1.82  \\
    IPW \& Wrong model imputation w/ MSI   & .00 & 1.95 & .00 & 1.99 & .00 & 2.13 \\ 

    \multicolumn{7}{l}{\underline{$N=500$, $10\%$ missing X}} \\
    Complete outcome w/o cov. adj. & .34 &     1.06  & .32 & 1.04 & .29 & 1.04  \\
    IPW w/o cov. adj. & .01  & 1.00  & 0.00 & 1.00   & .00 & 1.00 \\
    IPW \& Full X  & .01 & 3.01 & .00 & 3.16 & .00 & 3.09 \\
    IPW \& Complete covariate & .01 & 1.14 & .00 & 1.14 & .00 & 1.14 \\
    IPW \& Mean imputation w/o MSI  & .00 &     2.54 & .00 & 2.64 & .00 & 2.32  \\ 
    IPW \& Mean imputation w/ MSI  & .00 & 2.55   & .00 & 2.67 & .00 & 2.63 \\ 
    IPW \& Correct model imputation w/o MSI  & .00  & 2.57 & .00 & 2.70 & .00 & 2.39 \\ 
    IPW \& Correct model imputation w/ MSI  & .00 & 2.58 & .00 & 2.70  & .00 & 2.66 \\
    IPW \& Wrong model imputation w/o MSI   & .00  & 2.54 & .00 & 2.62 & .00 & 2.32 \\
    IPW \& Wrong model imputation w/ MSI   & .00  & 2.54 & .00 & 2.65 & .00 & 2.62 \\ 
    \bottomrule
    \end{tabular}
    }
    \caption{Simulation results with missing covariates and missing continuous outcomes (N=500, 5000 iterations). RE stands for relative efficiency compared to the IPW estimator without covariate adjustment; MSI stands for missing indicators.}
    \label{cont500-mis-y}
\end{table}

\section{Application} \label{sec:application}

The Childhood Adenotonsillectomy Trial (CHAT) \citep{marcus2013randomized,zhang2018national} is a multi-center, single-blind, randomized, controlled trial designed to test whether children, ages 5 to 9.9 years, with mild to moderate obstructive sleep apnea randomized to early adenotonsillectomy (eAT, treatment) will show better neurocognitive functioning than children randomized to watchful waiting with supportive care (WWSC, control). Children were randomized with a 1:1 ratio to both arms. Evaluations of participants were conducted at baseline as well as at 7 month post-baseline. This paper focuses on the change of the score from the Behavior Rating Inventory of Executive Function (BRIEF). Lower BRIEF scores indicate better executive functioning. An unadjusted analysis based on data from 392 children who had valid BRIEF measurements (195 and 197 in the eAT and WWSC arm, respectively) reveals that the average decrease in BRIEF score of the eAT arm exceeds that of the WWSC arm by 3.71 points and the difference is statistically significant ($ p < .001$). In this section, we first investigate whether adjusting for baseline covariates can improve the efficiency of the treatment effect estimation based on patients with a complete BRIEF score. We further apply our proposed weighting estimator (\ref{eq:Hajek}) to not only adjust for baseline covariates but also account for 61 patients without a valid BRIEF score. \\

\begin{table}[H]
  \centering
  \resizebox{\columnwidth}{!}{
    \begin{tabular}{lrrrrrr}
    \toprule
    \toprule
          \multicolumn{1}{l}{Participants with valid BRIEF score} & \multicolumn{1}{l}{Overall} &  \multicolumn{1}{l}{eAT group} & \multicolumn{1}{l}{WWSC group} &Unadjusted
         & \multicolumn{1}{l}{Missing } & \multicolumn{1}{l}{Corr. with outcome} \\
          & \multicolumn{1}{l}{$N=392$} &   \multicolumn{1}{l}{$N_1= 195$} & \multicolumn{1}{l}{$N_0= 197$}     &ASD &(\%) & \multicolumn{1}{l}{(p value)} \\
          \midrule
     \multicolumn{7}{l}{\textbf{Baseline categorical covariates (count) }} \\
    Gender (male) &  193     &    88   & 105 & .16 & 0     & -.08 (.11)    \\
    Race (black) &    207     &    102   &  105 & .02 & 0     & .05 (.33)    \\
    Tonsil size (0-50\%) &  98     & 44       & 54   & .11  & .8   &   .05 (.33)  \\
    Family income ($\leq 40000$) & 154  &  73     &  81  & .04 & 14  & -.03 (.61) \\
    Second-hand smoking (yes) & 123 & 61      &  62  & $<$.01 & 0     & .06 (.22)\\
    History of asthma (yes) & 126 & 61 &  65 & .03 & 2.3   & -.06 (.26)\\
    History of allergy (yes) &    165   &  81  & 84   & .03 & .3   & .01 (.89)\\
    History of eczema (yes) &  105          &     53  &  52   & .02 & .8   & -.02 (.70)\\
    History of prematurity (yes) & 60  &  27     &33   & .09 & 1.3   & .05 (.29)  \\
    \midrule
    \multicolumn{7}{l}{\textbf{Baseline continuous covariates (mean (sd))}} \\
    Age (years) &   6.6 (1.4)     &  6.6 (1.4)     &  6.6 (1.4)    & $<$.01 & 0     &  -.10 (.06)   \\
    BMI z-score &  .9 (1.3)     &   .9(1.4) & .9 (1.3)    & .03 & 3.3   &    -.13 (.01)     \\
    Birth weight (oz) &  109 (24.5)       & 110 (23.5)       &  108 (25.4)  & .10 & 6.9   &  -.03 (.55)  \\
    PSQ score &   .5 (.2)     &   .5 (.2)    &0.5 (.2)   & .02 & .8   &    -.19 ($<$.01)  \\
    OSA score &   53.7 (18.6)       & 53.4 (17.9)      & 54.0 (19.4)  & .03 & .5   &   -.23 ($<$.01) \\
    \midrule
    \multicolumn{7}{l}{\textbf{BRIEF Score}} \\
    Baseline & 50.1 (11.4) & 50.1 (11.2) & 50.1 (11.5) \\
    Endline & 48.7 (11.9 & 46.8 (11.6) & 50.5 (11.9) \\
    Change & -1.46 (8.8) & -3.33 (8.5) & 0.39 (8.8)\\ 
    \bottomrule
    \toprule
          \multicolumn{1}{l}{All Participants} & \multicolumn{1}{l}{Overall} &  \multicolumn{1}{l}{eAT group} & \multicolumn{1}{l}{WWSC group} &Unadjusted
         & \multicolumn{1}{l}{Missing } & \multicolumn{1}{l}{Corr. with outcome} \\
          & \multicolumn{1}{l}{$N=453$} &   \multicolumn{1}{l}{$N_1= 227$} & \multicolumn{1}{l}{$N_0= 226$}     &ASD &(\%) & \multicolumn{1}{l}{(p value)} \\
          \midrule
     \multicolumn{7}{l}{\textbf{Baseline categorical covariates (count) }}  \\
    Gender (male) &  219   & 101 &    118  &  .15 & 0     &  -.08 (.11)   \\
    Race (black) &    249  &  126 &    123  & .03 & 0 &  .05 (.33)   \\
    Tonsil size (0-50\%) &  113      & 50 & 63   & .13  & .9   &  .05 (.33)  \\
    Family income ($\leq40000$) & 182    &  90 &  92  & .02 & 14.1  &  -.02 (.72) \\
    Second-hand smoking (yes) & 142         & 72    &  70  & .02 & 0  & .06 (.22) \\
    History of asthma (yes) & 143          & 69 &  74  & .05 & 2.2  & -.06 (.26) \\
    History of allergy (yes) &    186         &  90     & 96   &.05   & .2   & .00 (.89) \\
    History of eczema (yes) &  123   &     62  &  61   & .02  & 1.1   & -.02 (.08)  \\
    History of prematurity (yes) & 68          &  32    & 36   & .05  & 1.1   & .05 (.29) \\
    \midrule
    \multicolumn{7}{l}{\textbf{Baseline continuous covariates (mean (sd))}} \\
    Age (years) &   6.6 (1.4) &  6.6 (1.4) &  6.5 (1.4)        & .05 & 0     &   -.10 (.06) \\
    BMI z-score &  .9 (1.3) & .9(1.4) &   .9 (1.2)     & .05 & 2.9   &  -.13 (.01)    \\
    Birth weight (oz) & 108 (24)      &  109 (22.9) & 107 (25.1)        & .10 & 6.6   &  -.03 (.05) \\
    PSQ score &   .5 (.2)  & .5 (.2) &   .5 (.2) & .06 & 1.3   & -.19 ($<$.01)  \\
    OSA score &   53.6 (18.6)  & 53.1 (18.3) & 54.1 (18.8)       & .05 & .4   & -.22 ($<$.01)   \\
    \midrule
    \multicolumn{7}{l}{\textbf{BRIEF Score}} \\
    Baseline & 49.9 (11.3) & 49.9 (11.1) & 50 (11.5)  & & .7& \\
    Endline & 48.7 (11.9) & 46.8 (11.6) & 50.6 (11.8) & & 12.8 & \\
    Change & -1.46 (8.8) & -3.33 (8.5) & 0.39 (8.8) & & 13.5 &  \\ 
    \bottomrule
    \end{tabular}%
    }
     \caption{Baseline characteristics of the CHAT study by treatment groups. The unadjusted ASD and correlation with outcome are computed ignoring missing values.}
  \label{baseline}%
\end{table}%

\indent Table \ref{baseline} displays a summary of the baseline demographic and clinical characteristics based on participants with a valid BRIEF score (392 patients) and all participants (453 patients). For each variable, we present the unadjusted absolute standardized difference (ASD) as a measure of baseline imbalance: $$\mbox{ASD} = \bigg|\sum_{i=1}^NX_{ij}Z_i/\sum_{i=1}^N Z_i- \sum_{i=1}^NX_{ij}(1-Z_i)/\sum_{i=1}^N (1-Z_i)\bigg|\bigg/\sqrt{[\widehat{\mbox{Var}}(X_{ij}|Z_i=1) +\widehat{\mbox{Var}}(X_{ij}|Z_i=0)]/2},$$
where $\widehat{\mbox{Var}}(\cdot)$ denote the sample variance $s^2$ for a continuous variable, and $\hat{p}(1-\hat{p})$ for a binary variable with empirical prevalence $\hat{p}$. 
Two binary covariates, gender and tonsil size, have ASD larger than $10\%$, which has been regarded as a common threshold for imbalance \citep{Austin2015}. In addition, univariate analyses reveal that the baseline BMI z-score, total score from the Pediatric Sleep Questionnaire (PSQ score) and total score from the Obstructive Sleep Apnea Quality of Life Survey (OSA score) are significantly correlated with the outcome ($p<0.05$).  The two covariates with baseline imbalance (gender and tonsil size), together with the three that are significantly correlated with the outcome (BMI z-score, PSA score and OSA score), as well as baseline BRIEF score, are subsequently included in the covariate adjustment analyses. These covariates exhibit mild missingness ranging from $0.4\%$ to $3.3\%$, except for gender, which is observed for all participants. 
\\ 
\indent We perform both mean imputation and model-based imputation for the missing covariates. For model-based imputation, we fit linear regression models with all the fully-observed covariates as predictors. We then conduct covariate adjustment using ANCOVA and OW with the 6 selected covariates based on participants with valid BRIEF measurements. The ANCOVA models include main effects of treatment, centered covariates, and centered missing indicators of those covariates with missingness, together with treatment-covariate interactions and treatment-missing-indicator interactions. Correspondingly, the treatment propensity score models contain the covariates and their missing indicators. To account for potential informative missing outcomes, we apply our proposed weighting estimator \ref{eq:Hajek} using the same treatment propensity score model. We use LASSO for covariate selection of the propensity score model for outcome missingness, where race, treatment-race interaction, and missing indicators for history of eczema, BMI z-score, PSQ score, and baseline BRIEF scores, are selected for the final model. 

We contrast the bootstrap variance estimates resulted from the covariate adjustment analyses to that from the unadjusted analysis. The results are presented in Table \ref{application-results}. The relative efficiency is computed as the ratio of the bootstrap variance of the unadjusted estimate to those of the adjusted ones. Overall, different combinations of imputation methods and adjustment methods produce similar treatment effect estimates, and on average adjusting for these three covariates leads to a 6\%-7\% of efficiency improvement over the unadjusted analysis. Inclusion of missing indicators in the treatment propensity score models does not appear to impact efficiency gain, implying that the covariate missingness is likely to fall under the MCAR or MAR setting. Table \ref{baseline} demonstrates that the covariate distributions for participants with a valid BRIEF score and all participants are very similar. This might explain that outcome missingness is not informative, yielding similar treatment effect estimates regardless of adjustment for the missing outcomes.

\begin{table}[H]
    \centering
    \resizebox{\columnwidth}{!}{
    \begin{tabular}{clcccc}
    \toprule
    \toprule
    \multicolumn{6}{c}{\underline{Results based on patients with complete outcomes}} \\
           Covariates  & \multicolumn{2}{c}{Method}  & $\hat\tau$   & $\hat{V}_{boot}$ & Rel.Eff.$_{boot}$ \\
    \midrule
    None & \multicolumn{2}{l}{Unadjusted } &  -3.71 & 0.86 & 1.00 \\
    \midrule
         Gender, Tonsil size, baseline BRIEF, &\multirow{2}*{Mean imp.}& ANCOVA  &  -3.69 & 0.82 & 1.06 \\
        BMI z-score, PSQ score, OSA score & & OW& -3.69 & 0.81 & 1.07 \\
        w/o missing indicators &\multirow{2}*{Model-based imp.}& ANCOVA  & -3.69 & 0.82 & 1.06 \\
        & & OW & -3.69 & 0.81 & 1.07 \\
        \midrule
        Gender, Tonsil size, baseline BRIEF, &\multirow{2}*{Mean imp.}& ANCOVA  & -3.70 & 0.82 & 1.06 \\
        BMI z-score, PSQ score, OSA score & & OW& -3.71 & 0.81 & 1.06 \\
        w/ missing indicators &\multirow{2}*{Model-based imp.}& ANCOVA  & -3.70 & 0.82 & 1.06 \\
        & & OW & -3.71 & 0.81 & 1.07 \\
        \midrule
        \midrule
        \multicolumn{6}{c}{\underline{Results based on all patients}} \\
           Covariates  & \multicolumn{2}{c}{Method}  & $\hat\tau$   & $\hat{V}_{boot}$ & Rel.Eff.$_{boot}$ \\
    \midrule
    None & \multicolumn{2}{l}{IPW w/o cov. adj. } & -3.64 & 0.88 & 1.00 \\
    \midrule
         Gender, Tonsil size, baseline BRIEF, BMI z-score, & Mean imp. & \multirow{2}*{IPW + OW} & -3.67 & 0.82 & 1.07 \\
         PSQ score, OSA score, w/o missing indicators & Model-based imp. &  & -3.67 & 0.82 & 1.07 \\
        \midrule
        Gender, Tonsil size, baseline BRIEF, BMI z-score, & Mean imp. & \multirow{2}*{IPW + OW} & -3.70 & 0.82 & 1.06 \\
        PSQ score, OSA score, w/ missing indicators & Model-based imp. & & -3.70 & 0.82 & 1.06 \\
        \bottomrule
    \end{tabular}
    }
    \caption{Modeling results}
    \label{application-results}
\end{table}

\begin{figure}[H]
    \centering
    \subfloat[\centering MCAR]{{\includegraphics[scale=0.3]{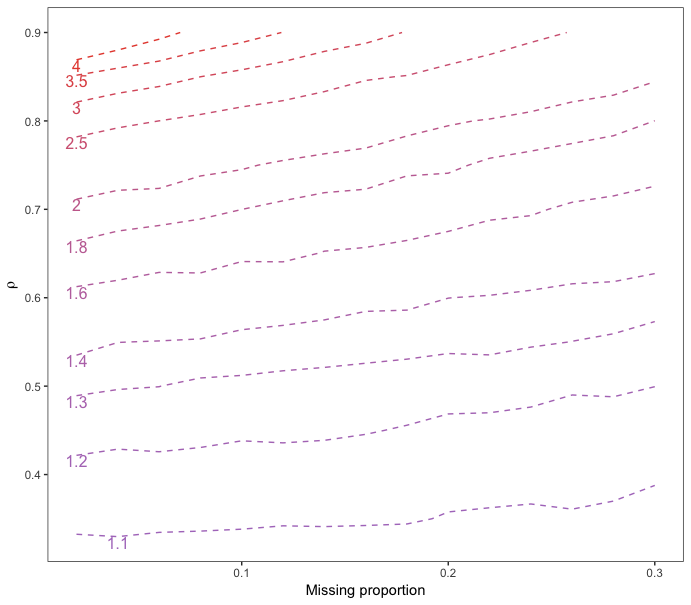} }}%
    \qquad
    \subfloat[\centering MNAR]{{\includegraphics[scale=0.3]{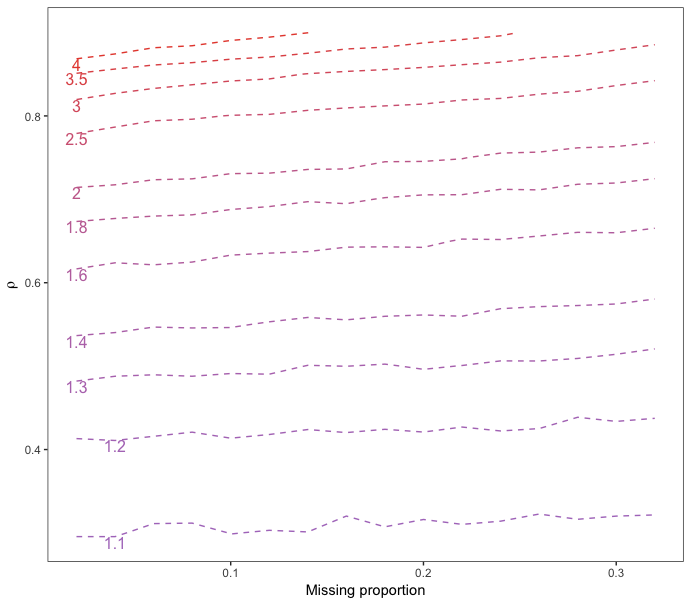} }}%
    \caption{Relative efficiency of ANCOVA with missing indicators after mean imputation to the unadjusted analyses under MCAR and MNAR mechanisms with varying values of $\rho$ and missing proportions}
    \label{eff}
\end{figure}

\indent Through the use of a pseudo variable, we further investigate how a covariate's correlation with the outcome and its level of missingness relate to the amount of possible efficiency gain from adjusting for that covariate. Based on participants with a valid BRIEF score, we generate $X_i = \rho\hat{s}_i + \sqrt{(1-\rho^2)\hat{v}_s} \varepsilon_i$, where $
\varepsilon_i\overset{iid}{\sim} N(0,1)$, $\hat{s}_i$'s are the residuals from fitting the unadjusted model $Y\sim Z$ and $\hat{v}_s$ is the sample variance of $\hat{s}_i$. The correlation between the $X_i$'s generated in this way and $Y_i$ is approximately $\rho$. By removing the effect of $Z_i$, $X_i$ is uncorrelated with the randomization indicator $Z_i$.   We then consider missingness of the variable $X$ following either a MCAR mechanism or a MNAR one where $\mbox{logit}\{Pr(R_{i}^{x}=1)\} = \delta_0 + X_i$. The value of $\delta_0$ is varied to achieve specific levels of missingness. The empirical efficiency of the treatment effect estimates resulting from ANCOVA with missing indicators after mean imputation of missing values in $X$ over repeated experiments is then compared to that of the unadjusted estimators. Fig. \ref{eff} shows the contour plot of relative efficiency for varying levels of $\rho$ and proportions of missingness under MCAR and MNAR. The relative efficiencies are computed as the ratios of the average bootstrap variances of the estimators from covariate adjustment analyses to that from the unadjusted analysis. From the plot for MCAR, we observe that in order to gain a 30\% increase in efficiency, the covariate needs to have a correlation of approximately 0.5 or higher with the outcome. The efficiency gain diminishes with increased percentages of missingness, which is expected.  The plot for the MNAR case is largely similar.

\section{Discussion} \label{sec:discussion}
We investigate covariate adjustment in randomized controlled trials with missing data in both baseline covariates and outcomes. When there are missing data in covariates only, performing covariate adjustment on the imputed data generally reduces the variance of treatment effect estimate. To gain additional efficiency, it is important to add the missingness indicator in the covariate adjustment model, but how the missing data is imputed is not important. When both covariates and outcomes are partially observed, we specify MAR assumptions for the missing outcomes, where the outcome missingness can depend on the fully-observed pre-treatment covariates and treatment indicators,  the missingness indicators of the partially observed covariates, and the values of the partially observed covariates when observed.  Under these assumptions, as long as the propensity model for the probability of $Y$ being observed includes the missingness indicator of $X$ and its interaction with $X$ (equivalent to imputing the missing values by zero) as the predictors, it does not matter how the missing covariate values are imputed. Based on this result, we propose two methods, one via outcome regression and one via weighting, with specific models for covariate adjustment. 

Our simulations confirm the analytical results. Regardless of the missingness mechanism for the partially observed covariate, we observe improvement in the precision of the treatment effect estimate when performing covariate adjustment based on the imputed data. When covariates are missing not at random, including the covariate missingness indicators in the adjustment model could lead to further efficiency gain. The magnitude of improvement depends on the proportion of missingness and the strength of the association between the adjusted covariate and the outcome. Simulation results also suggest that the outcome regression and overlap weighting method lead to similar results in many settings. Therefore, we conjecture that the asymptotic equivalence between the outcome regression and propensity score weighting adjustment methods with complete covariates extends to the case of incomplete data. Nonetheless, the overlap weighting method is arguably more stable and easier to implement than outcome regression for non-continuous outcomes. As a general recommendation, in the presence of incomplete data in randomized trials, we recommend analysts to impute all the missing covariate values with mean (or median for categorical variables), include the covariate missingness indicators and their interaction with $X$ in either the treatment propensity score or outcome models, and perform a specific covariate adjustment method (either outcome regression or propensity score weighting) on the imputed data.

The use of missing indicator method in the context of observational studies has been investigated extensively \citep{jones1996indicator,greenland1995critical,song2021missing}. However, randomized experiments are distinct from observational studies. Heuristically, randomization balances all measured and unmeasured covariates, including the missingness patterns in the baseline covariates, between treatment and control units in large samples. Thus adding the missingness indicators in the outcome regression leads to consistent estimator of the ATE and could improve efficiency when the missingness patterns are predictive of the outcome, regardless of the specific imputed values. This no longer stands in observational studies and we expect that the imputation data quality affects final estimates of the treatment effects.    

In randomized clinical trials, stratified randomization is commonly used to avoid potential imbalance on important baseline covariates during the randomization stage.   When stratified randomization is used, because the randomization is based on study participants' stratification factors, these variables are typically fully observed. After implementing stratified randomization, the recommended analysis approach is to adjust the baseline covariates used in the stratified randomization process \citep{kahan2012improper, wang2021model}. To further improve precision of treatment effect estimation, one can additionally adjust for baseline prognostic variables not used for stratification  \citep{zhang2021discussion}. The analysis methods for covariate adjustment in the presence of missing data we investigate in this paper can be applied to studies that utilize simple randomization or stratified randomization; in the latter case, the stratification factors and additional prognostic variables that are fully observed can be included in the set of covariates $V$. The partially-observed prognostic baseline variables can be included in the set of covariates $X$. 

We compare covariate adjustment methods for partially-observed covariates ($X$) based on outcome regression and propensity score weighing in terms of efficiency gain in randomized studies. Various doubly robust (DR) methods that combine a propensity score model and an outcome model have been proposed \citep{robins1994estimation,robins1995analysis,lipsitz1999weighted,lunceford2004stratification,neugebauer2005prefer,bang2005doubly}. In causal inference with observational data, a DR estimator can be constructed by combining a propensity score model for the treatment assignment mechanism and a model for the distribution of the counterfactual data.  When the outcome $Y$ is missing at random, a DR estimator can be constructed combining a propensity score model for the missingness mechanism and a model for the distribution of the complete data. In both cases, the resulting doubly robust estimators remain consistent when either model is correctly specified.   Augmented inverse probability weighted estimators that account for both confounding and missing data have also been investigated \citep{williamson2012doubly}. In randomized studies with fully observed outcome data $Y$, because the treatment propensity score model is always correct (the true propensity is a constant for all units),  the additional outcome model becomes unnecessary \citep{zeng2021propensity} When the outcome $Y$ is missing at random, it would be useful to investigate how to construct DR estimators in the presence of missing covariates in future research.

In this paper, we focus on continuous and binary outcomes, but time-to-event outcomes with right censoring is common in randomized trials. For the setting of missing covariate values in the Cox models for a survival outcome, White and Royston compared the use of several imputation models and recommended including the event indicator and the Nelson-Aalen estimator of the cumulative baseline hazard in the imputation model \citep{white2009imputing}.  Yi et al. considered Cox regression with survival-time-dependent missing covariate values, proposed an inverse propensity weighting method with the propensity estimated by nonparametric kernel regression, and pointed out the need for further investigation of survival-time-dependent time-varying missingness propensity estimation \citep{yi2020cox}. However, there lacks investigation on covariate adjustment with missing data both in covariates and survival outcomes, an important direction for future research.    
Another direction is to investigate the implication of missing data when performing covariate adjustment in cluster randomized trials, where the unit of randomization is a cluster (such as hospitals) and outcomes and covariates are measured at individual level \citep{turner2017a}. Systematic reviews suggested that the median number of clusters in a CRT is around 20 \citep{murray2008design}. Such small number of clusters renders covariate imbalance prevalent in CRT. When data are fully observed, an augmented generalized estimating equations approach has been proposed to improve efficiency and validity of estimation in cluster randomized trials leveraging cluster-level and individual-level covariates \citep{stephens2012augmented}. Doubly robust estimators for the marginal treatment effect have also been proposed in the presence of informative missing outcome data \citep{prague2016accounting}.  It is critical to differentiate imbalance caused by random chance and by post-randomization selection: the former can be addressed by covariate adjustment but the latter can not and would require additional data and assumptions \citep{li2021clarifying}. 

\clearpage

\section*{Data Availability Statement} 
The Childhood Adenotonsillectomy Trial data are available upon reasonable request at https://sleepdata.org. 

\section*{Acknowledgement}
We thank the Editor, Associate Editor, and reviewers for their helpful comments, which improved the article. We thank Dr. Peng Ding and Dr. Anqi Zhao for constructive comments. Chang and Wang were partly supported by grant R01 AI136947 from the National Institute of Allergy and Infectious Disease. Li was partly supported by PCORI contract ME-2019C1-16146. The Childhood Adenotonsillectomy Trial (CHAT) was supported by the National Institutes of Health (HL083075, HL083129, UL1-RR-024134, UL1 RR024989). The National Sleep Research Resource was supported by the National Heart, Lung, and Blood Institute (R24 HL114473, 75N92019R002). The contents of this article are solely the responsibility of the authors and do not necessarily represent the view of NIH or PCORI.

\bibliographystyle{plainnat}
\bibliography{RCT_missingcov}

\end{document}